\documentclass[10pt,aps,amsmath,showpacs,amssymb,longbibliography,nofootinbib,pre,twocolumn,epsfig]{revtex4}
\usepackage{bmpsize}
\usepackage[dvips,usenames]{color}
\usepackage{amsmath}
\newcommand{\av}[1]{\langle {#1} \rangle}

\usepackage{graphicx}
\usepackage{lipsum}

\begin{document}

\title{Heterogeneous excitable systems exhibit Griffiths phases below 
hybrid phase transitions} 

\author{G\'eza \'Odor (1) and Beatriz de Simoni (2)}
\affiliation{(1) Institute of Technical Physics and Materials Science,
Center for Energy Research, P. O. Box 49, H-1525 Budapest, Hungary \\ 
(2) Budapest University of Technology and Economics, M\H uegyetem rkp. 3., 
H-1111 Budapest, Hungary}


\date{\today}


\begin{abstract}


In $d > 2$ dimensional, homogeneous threshold models discontinuous transition 
occur, but the mean-field solution provides $1/t$ power-law activity decay 
and other power-laws, thus it is called mixed-order or hybrid type. 
It has recently been shown that the introduction of quenched disorder rounds the 
discontinuity and second order phase transition and Griffiths phases appear. 
Here we provide numerical evidence, that even in case of high graph dimensional 
hierarchical modular networks a Griffiths phase in the $K=2$ threshold model
is present below the hybrid phase transition.
This is due to the fragmentation of the activity propagation by modules, 
which are connected via single links.
This provides a widespread mechanism in case of threshold type of heterogeneous 
systems, modeling the brain or epidemics for the occurrence of dynamical 
criticality in extended Griffiths phase parameter spaces. 
We investigate this in synthetic
modular networks with and without inhibitory links as well as in the
presence of refractory states.

\end{abstract}

\maketitle


\section{Introduction}


Phase transitions in genuine nonequilibrium systems have often been 
investigated among reaction-diffusion (RD) type of models exhibiting 
absorbing states \cite{marro2005,HHL}. 
In many cases mapping to surface growth, spin systems 
or stochastic cellular automata have been used. Criticality allows us to define 
universality classes, defined by the scaling exponents, which have been 
explored in homogeneous systems~\cite{rmp,odorbook}. 
In heterogeneous network models the situation is less clear. 
Hybrid phase transition (HPT) means that at the transition point the 
order parameter exhibits a jump, in conjunction with critical phenomena 
related to it. It can mean avalanches of activity at the transition 
point with power-law (PL) size distribution for example. 
Such type of transitions have been known for a long time~\cite{Cardy_1981}, 
for example at tricriticality~\cite{PhysRevE.70.026114,Gras-av}, but had 
not been the focus of research and the term appeared later.
HPT-s have been found in network science in case of 
$k$-cores ~\cite{PhysRevLett.96.040601}, interdependent 
networks~\cite{PhysRevE.93.042109} and multiplexes~\cite{PhysRevE.89.042801}. 

The "mixed-order" naming for the same phenomena in statistical physics 
arouse some years ago~\cite{PhysRevLett.112.015701} by the exactly soluble 
one-dimensional Ising model with long range interactions.
It is also known to appear in nonequilibrium models, exhibiting transition
to absorbing states~\cite{PhysRevE.95.022109}. Further examples include 
critical models at extended surface 
defects~\cite{PhysRevE.95.010102,PhysRevE.97.012111} and 
synchronization~\cite{PhysRevE.72.046211,PhysRevLett.106.128701,PhysRevE.87.032106}.

Criticality is an ubiquitous phenomenon in nature as systems can benefit
many ways from it. As correlations and fluctuations diverge~\cite{PhysRevLett.110.178101}
in neural systems working memory and long-connections can be generated 
spontaneously~\cite{Johnson} and the sensitivity to external signals is maximal. 
Furthermore, it has also been shown that information-processing capabilities 
are optimal near the critical point. 
Therefore, systems tune themselves close to 
criticality via self-organization (SOC)~\cite{SOC,Chialvo2010}, presumably
slightly below to avoid blowing over excitation.
Besides, if quenched heterogeneity (that is called disorder compared
to homogeneous system) is present, rare-region (RR) effects~\cite{Vojta2006b}
and an extended semi-critical region, known as Griffiths Phase 
(GP)~\cite{Griffiths} can emerge. 
RR-s are very slowly relaxing domains, remaining in the opposite phase than
the whole system for a long time, causing slow evolution of the order parameter.
In the entire GP, which is an extended control parameter region around the
critical point, susceptibility diverges and auto-correlations exhibit fat tailed,
power-law behavior, resulting in bursty behavior~\cite{burstcikk}, frequently observed
in nature~\cite{Karsai_2018}.
Even in infinite dimensional systems, where mean-field behavior is expected, 
Griffiths effects~\cite{Cota2016} can occur in finite time windows. 

It is known that strong disorder can round or smear phase transitions~\cite{Vojta2006b}.
According to the arguments by Imry-Ma ~\cite{IM} and 
Aizenman-Wehr~\cite{Aw}, first-order transitions do not exist in 
low-dimensional disordered systems. It has recently been shown~\cite{round,round2} 
that this is true in genuinely nonequilibrium systems~\cite{marro2005,odorbook}.

Experimental and theoretical research provide evidence that
the brain operates in a critical state between sustained
activity and an inactive 
phase~\cite{BP03,T10,H10,R10,PhysRevLett.110.178101,PONCEALVAREZ20181446,MArep}.
Criticality in general occurs at continuous, second order phase transitions.
On the other hand, meta-stability and hysteresis are also common in the 
brain behavior. They are related to the ability to sustain stimulus-selective 
persistent activity for working memory~\cite{bistab-exp}.
The brain rapidly switches from one state to another in response to
stimulus, and it may remain in the same state for a long time after 
the end of the stimulus. It suggests the existence of a repertoire 
of meta-stable states. There have been several model describing this
~\cite{doi:10.1162/neco.2007.19.11.3011,PhysRevE.97.062305}.
It introduces an apparent contradiction, because meta-stability and hysteresis 
occur in general at first order, discontinuous phase transitions.
But the brain can operate at different regimes close to the critical point 
which can provide the desired advantages for biological systems. 
Another possible resolution for the above controversy is the operation 
at a transition of hybrid type. 
It has also been suggested in a recent theoretical 
work~\cite{buendia2020hybridtype}.

Threshold type of systems, like the integrate and fire models of
the brain~\cite{KC}, are also suggested to describe other phenomena, 
like power-grids~\cite{Car2,SWTL18,POWcikk}, crack and fracture 
formation~\cite{Alava_2006}, contagion~\cite{Th-cont}, etc. 
In these models HPT can emerge naturally, thus the present results can 
also be relevant.

Heterogeneity effects are very common in nature and result in dynamical 
criticality in extended GP-s, in case of quasi static quenched disorder 
approximation~\cite{Munoz2010}. 
This leads to avalanche size and time distributions, with non-universal 
PL tails. It has been shown within the framework of modular 
networks~\cite{Munoz2010,HMNcikk,Cota_2018} and a large human connectome 
graph~\cite{CCcikk}. In this study we re-use the hierarchical modular
network of ~\cite{HMNcikk} and provide numerical evidence that above
the GP a HPT emerges. Meta-stable states and 
hysteresis behavior can also be found, thus this system can oscillate
between up and down states, depending on the level of oscillations,
without the need of oscillators at the nodes, as in case of
the Ginzburg-Landau theory, suggested to model cortical dynamics~\cite{MunPNAS}.
By extending our model we will show that the proposed mechanism is very 
general, providing an explanation for the observed
wide range of scale-free behavior below the transition point.


\section{Mean-field approximation}


Discrete threshold models can be defined as two-state systems: $x_i=0,1$
(inactive, active) at sites $i$, with a conditional activation rule, 
depending on the sum of activity of neighbors compared to the 
threshold value $K$:
\begin{equation}\label{thresh}
\sum_{j} x_j w_{i,j}  \ge K \ ,
\end{equation}
where $w_{i,j}$ is the weight of the link connecting site $j$ to $i$.
In interacting homogeneous systems $w_{i,j}$ is just the adjacency
matrix element: $A_{i,j}$, which is 1 if nodes are connected or 0
otherwise.
To describe stochasticity this activity creation can be accepted 
with probability $\lambda$, competing with an activity removal process
of probability $\nu$.
The mean-field description of the threshold model of $N$ nodes can be obtained 
in a similar way as in case of RD systems~\cite{PhysRevE.67.056114}.
That work is defined on the lattice, but we can apply it for other graphs.
In~\cite{PhysRevE.67.056114} it was shown that discontinuous jump occurs in 
mean-field models of the $n > m$ RD systems, in which $n$ neighboring 
particles are needed for creation and $m$ neighbors for removal. 
Here we don't have diffusion and particles, but the activity can be 
considered as site occupancy and we can map the threshold model with $K=2$ 
to an RD system with $n = 2$ active neighbors necessary for creation and 
$m = 1$ for spontaneous removal. In the presence of inhibitions
$n > 2$ is needed for creation at nodes with negatively weighted incoming links
which increases the inactive phase.

In the mean-field approximation the probability of site activity is $\rho$, 
and two active neighboring sites can occur in a $(N-1)(N-2)/2$ way thus
the creation rate in case of a global acceptance probability $\Lambda$ is
\begin{equation}
\frac{1}{2} (N-1)(N-2)\Lambda \rho^2 (1-\rho) \ .
\end{equation}
Let us call : $\Lambda (N-1)(N-2)/2 = \lambda$. 
For a full graph of $N$ nodes we can setup the rate equation
\begin{equation}
    \frac{d \rho } {d t} = \lambda \rho^2 (1-\rho) - \nu\rho \ ,
\end{equation}
which in the $N\to\infty$ limit provides $\lambda_c=0$, but for finite graphs 
$\lambda_c > 0$.
In the steady state we have
\begin{equation}
    \lambda \rho^2 (1-\rho) - \nu\rho = 0 \ .
\end{equation}
By imposing the condition
\begin{equation}
    \nu = 1 - \lambda \ ,
\end{equation}
we obtain
\begin{equation}
    \lambda \rho (1-\rho) - (1-\lambda) = 0 \ ,
\end{equation}
which can be solved as
\begin{equation}
    \rho = \frac{ \lambda \pm \sqrt{ \lambda^2 - 4 \lambda (1-\lambda)}} {2 \lambda} \ .
\end{equation}
The solution is real and positive if
\begin{equation}
\lambda > \lambda_c = 4/5 \ ,
\end{equation}
providing a threshold within a system of size $N$
\begin{equation}
\Lambda_c = \frac{8}{ 5 (N-1) (N-2) } 
\end{equation}
and an order parameter for $\Lambda \to \Lambda_c^+$
\begin{equation}
\rho_c = 1/2 \ .
\end{equation}
It is important to realize that in the $N \to \infty$ limit $\Lambda_c\to 0$, 
thus there is no inactive phase in the thermodynamic limit. 
But as real systems are always finite sized, we can observe this
hybrid phase transition in them even in the mean-field limit.
Similar phenomenon has recently been reported in contagion 
models~\cite{PhysRevE.98.012311}. Furthermore, in case of the
presence of inhibitory couplings the HPT at finite transition
rate may survive the $N \to \infty$ limit in high dimensional 
systems.

At this transition point we can determine how the density approaches $\rho_c$:
\begin{equation}\label{mf-decay}
\rho(t) - \rho_c \sim t^{-1} \ . 
\end{equation}
Thus here we find PL dynamical behavior even though the transition is 
discontinuous, as in other known hybrid or mixed order transitions.
For $\lambda < \lambda_c$ we have $\rho=0$ stable solution and 
exponentially decaying activity.
Right above the transition the steady state density vanishes with a
square-root singularity as in case of k-core~\cite{PhysRevLett.96.040601} 
or multiplex percolation hybrid transitions~\cite{PhysRevE.89.042801}:
\begin{equation}
(\rho-\rho_c) \propto (\lambda-\lambda_c)^{1/2} \ ,
\end{equation}
but unlike the contact process~\cite{Harris74} near multiple 
junctions~\cite{PhysRevE.95.022109}, or the Kuramoto model with 
uniform frequency distribution~\cite{PhysRevE.72.046211},
which thus belong to another hybrid universality class.
In the following sections we investigate what happens to this 
HPT if we implement the threshold model on a hierarchical 
modular network (HMN).


\section{Threshold model on hierarchical modular networks}


In this section we describe the HMN models we use for the simulations.
It is important to note that we believe that hierarchy is not relevant,
but that modularity is what enhances RR effects as in case of the study~\cite{Cota_2018}.
The models are motivated by brain networks originated from Kaiser and Hilgetag,
who performed numerical studies to investigate the effects of different 
topologies on the activity spreading~\cite{KH}.
Their hierarchical model reflects general features of brain connectivity 
at large and mesoscopic scales, where the nodes were intended to 
represent cortical columns instead of individual neurons.
The connections between them were modeled excitatory, since there appears 
to be no long-distance, inhibitory connections within the cerebral 
cortex~\cite{LatNir2004}.

The network was generated beginning with the highest level and adding 
modules to the next lower level with random connectivity within modules. 
Kaiser and Hilgetag explored hierarchical networks with different numbers of hierarchical 
levels and numbers of sub-modules at each level. The total, average node 
degree was set to a fixed value, motivated by comparative experimental studies.
However, they investigated different topologies by varying the edge density 
across the levels. All the tested HMN-s were small-world type, 
i.e. exhibited infinite topological dimensions. 

The spreading model they investigated was a two-state threshold model, 
in which nodes became activated with probability $\lambda$, when at 
least $K$ of their directly connected node neighbors were active at the 
same time or spontaneously deactivated, with probability $\nu$. 
Note that this model is very similar to RD models known in statistical 
physics~\cite{rmp,odorbook}, with a synchronous cellular automaton (SCA) 
update. Without loss
of generality this algorithm produces faster dynamical scaling results 
for threshold models than those with random sequential updates.

In this paper we investigate versions of HMN-s, which possess increasing 
edge density from top to bottom levels. Clearly, such topologies can be 
expected to be more suitable for activity localization and for 
RR effects. 


\begin{figure}[h]
\includegraphics[height=6.5cm]{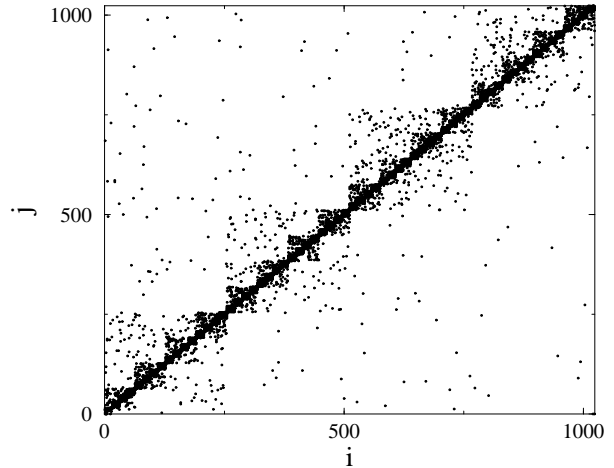}
\caption{Plot of the adjacency matrix of an $N=1024$ sized sample of the 
HMN2d graph used.
Black dots denote connections between nodes $i$ and $j$. The 4-level
structure is clearly visible by the blocks along the diagonal, additional
long-range edges are scattered points around it.} 
\label{Amat}
\end{figure}


One can also make a correspondence with the spatially embedded
networks~\cite{Barthelemy}. These networks have long links, with 
algebraically decaying probabilities in the Euclidean distance $R$ as
\begin{equation} 
p(R) \sim R^{-s}.
\label{BB}
\end{equation}
We added random long links by level-to-level from top to bottom, similar to 
in~\cite{HMNcikk}. 
The levels $l=0, 1, ...,l_{max}$ are numbered from bottom to top. 
The size of domains, i.e the number of nodes in a level, grows as 
$N_l = 4^{l+1}$ in the case of the $4$-module construction, related to 
a tiling of the 2d base lattice, due to the rough distance level relation:
\begin{equation} 
p_l = b (\frac{1}{2^s})^l .
\end{equation}
Here $b$ is related to the average degree $\langle k \rangle$ of nodes , which
was prescribed to be $\langle k \rangle= 12$ for this construction.

We connected nodes in a hierarchical modular way as if they were embedded 
in a regular two-dimensional lattice (HMN2d) as shown by the
adjacency matrix on Fig.~\ref{Amat}, similarly as in~\cite{HMNcikk}.
The $4$ nodes of the level $l=0$ were fully connected.
The single connectedness of the networks is guaranteed by additional 
linking of these $4$-node modules, by two edges to the subsequent ones: 
the first and the last nodes of module (i) to the first node of module (i+1). 
Accidental multiple connections were removed and self-connections 
were not allowed.
Note that the single connectedness at low level does not result in 
stable steady states in case of the threshold value $K=2$.

\begin{figure}[h]
\includegraphics[height=5.5cm]{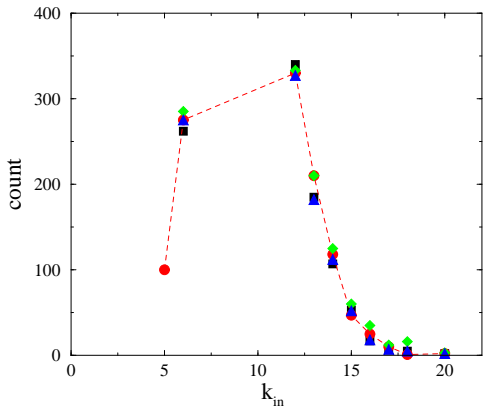}
\caption{In-degree distribution of 4 randomly selected $l=5$ HMN2d graphs.}
\label{indegfig}
\end{figure}

The in-degree distribution of 4 randomly selected graphs with $N=4096$ 
nodes can be seen on Fig.~\ref{indegfig}. 
The lowest in-degree is always $k^{in}_i \ge 5$.
The modularity quotient of the networks is high: $Q > 0.9$, defined by
\begin{equation}
Q=\frac{1}{N\av{k}}\sum\limits_{ij}\left(A_{ij}-
\frac{k_ik_j}{N\av{k}}\right)\delta(g_i,g_j),
\end{equation}
where $A_{ij}$ is the adjacency matrix and $\delta(i,j)$ is the Kronecker
delta function.
The Watts-Strogatz clustering coefficient \cite{WS98} of a
network of $N$ nodes is
\begin{equation}\label{Cws}
C = \frac1N \sum_i 2n_i / k_i(k_i-1) \ ,
\end{equation}
where $n_i$ denotes the number of direct edges interconnecting the
$k_i$ nearest neighbors of node $i$. $C=0.295$ is about $10$ times
higher than that of a random network of same size $C_r=0.0029$,
defined by $C_r = \langle k\rangle / N$.
The average shortest path length is defined as
\begin{equation}
L = \frac{1}{N (N-1)} \sum_{j\ne i} d(i,j) \ ,
\end{equation}
where $d(i,j)$ is the graph distance between vertices $i$ and $j$.
In case of this typical network $L = 6.74$, about twice
larger than that of the random network of same size: $L_r = 3.615$, 
following from the formula~\cite{Fron}:
\begin{equation}
L_r = \frac{\ln(N) - 0.5772}{\ln\langle k\rangle} + 1/2 \ .
\end{equation}
So this is a small-world network, according to the definition of
the coefficient~\cite{HumphriesGurney08}:
\begin{equation}
\sigma = \frac{C/C_r}{L/L_r} \ ,
\label{swcoef}
\end{equation}
because $\sigma=5.363$ is much larger than unity.

We estimated the effective topological dimension using the breadth-first 
search (BFS) algorithm: $d=4.18(5)$, defined by $N(r) \sim r^d$, where we 
counted the number of nodes $N(r)$ with chemical distance $r$ or less
from the seeds and calculated averages over the trials.
Note, that this is just an estimate for the finite sized graph,
because we know that $d\to\infty$ is expected for $s=3$.
It renders this model into the mean-field region, because 
for threshold models the upper-critical dimension is $d_c\le 4$.
Still, due to the heterogeneous structure, we find very non-trivial dynamical 
GP scaling behavior as will be shown in the following sections.


\section{Dynamical simulations}


Time dependent simulations were performed for single active seed initial
conditions. It means that a pair of nodes is activated at neighboring 
sites: $x_i=x_{i+1}=1$, in an otherwise fully inactive system. It can trigger an 
avalanche, a standard technique in statistical physics to investigate
critical initial slip~\cite{HHL}. We measured the spatio-temporal size
$s = \sum_{i=1}^N \sum_{t=1}^T x_i$ and the duration of the avalanches 
($T$) for tens of thousands of random initial conditions: both initial sites and 
initial graph configurations.
The graphs we investigated had $l=4,5,6,7$ levels, containing 
$N=1024, 4096, 16384, 65536$ nodes, respectively. The average node degree was 
$\langle k\rangle \simeq 12$, after the low level linking and the
removal of accidental multiple edges. The ratio of short and long links 
was $\simeq 0.6$. 

We have set $\nu=1-\lambda$ and updated the sites at discrete time steps, 
i.e. set the state variables $x'(i)=1$ if it was inactive $x(i)=0$ and
the sum of active neighbors $\sum_{j} x(j)$ exceeded $K=2$ with 
probability $\lambda$ or to $x'(i)=0$ with probability $1-\lambda$
if it was active $x(i)=1$. Following a full sweep of sites we 
wrote $x(i) = x'(i)$ for all nodes, corresponding to one Monte Carlo
step (MCs), throughout the study we measure time in MCs units.
We have measured the density of active nodes $\rho(t) = 1/N \sum_{i=1}^N x_i$.


\subsection{Excitatory model}

\begin{figure}[h]
\includegraphics[height=6.5cm]{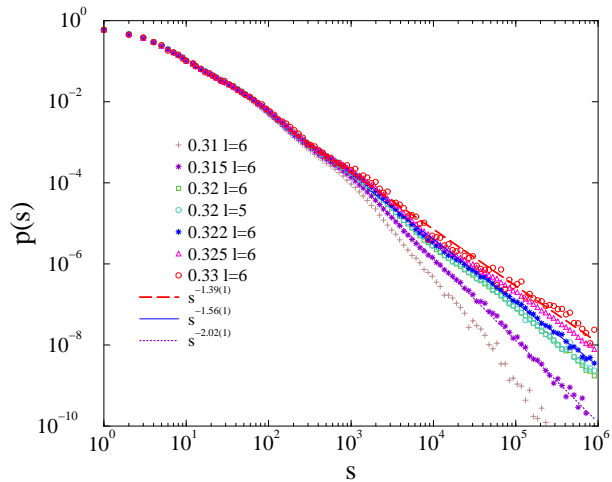}
\caption{Avalanche size distributions at different $\lambda$ branching rates,
denoted by the symbols, in the presence of excitatory links in the HMN2d
with $l=5,6$ levels. From top to bottom curves: 
$\lambda=$ 0.33, 0.325, 0.322, 0.32 ($l=5$ cyan and $l=6$ green), 0.315, 0.31. 
Dashed lines show PL fits for the tails: $s > 1000$
at $\lambda=$ 0.315, 0.322, 0.33.} 
\label{elo-hmn}
\end{figure}

The simulations were run for $T=10^7$ MCs, or until the system
goes to a fully inactive state, corresponding to the end of the
avalanche. We computed the probability density functions 
of avalanche sizes $p(s)$ and final survival time distributions $p(t)$.  
We repeated these simulations for different $\lambda$ branching rates, 
by increasing its value. As Fig.~\ref{elo-hmn} shows we don't see
exponential decays as should be in the inactive phase of a mean-field
model. Instead, there are PL-like tails for $\lambda > 0.31$, modulated slightly
by oscillations, which is a well known phenomenon when discrete spatial
periodicity is present, here the size of the modules.
The slopes of the PL tails vary from $\tau=2.02$ to $\tau=1.39$ as we 
increase $\lambda$ from $0.315$ to $0.33$.

\begin{figure}[h]
\includegraphics[height=6.5cm]{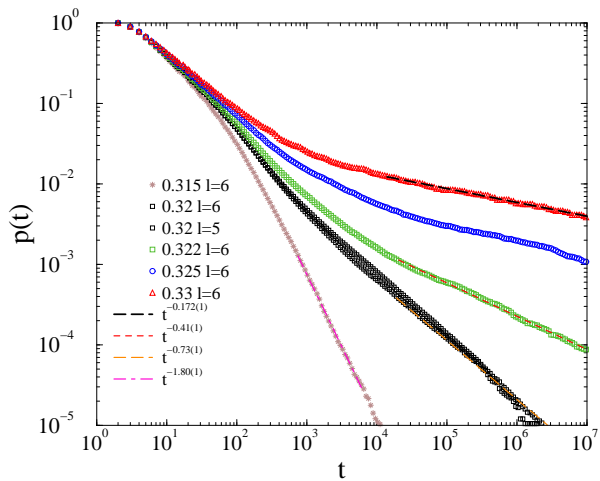}
\caption{\label{P21seed}Survival probability of the activity at different
branching rates in the $K=2$ threshold model with excitatory links. 
From top to bottom curves: $\lambda=$ 0.33, 0.325, 0.322, 0.32 
($l=5$ and $l=6$), 0.315. 
Dashed lines show PL fits for the tails: $s>10^4$
at $\lambda=$ 0.315, 0.322, 0.33.} 
\end{figure}

Non-universal PL tails are more clearly visible on the avalanche survival time 
distributions plotted on the graph, shown on Fig.~\ref{P21seed}. 
Here we can observe a greater variation as moving from $\lambda = 0.315$ 
with $\delta=1.80(1)$ to $\lambda = 0.33$ with $\delta=0.172(1)$. 
The avalanche duration distributions can be deduced from these curves as 
the time integral, thus $\delta$ is related to the duration exponent of 
$P(t) \propto t^{-\tau_t} $ as
\begin{equation}\label{taut-del}
\tau_t = 1 + \delta \ .
\end{equation}
These non-universal PL-s suggest that Griffiths effects are present, as 
reported in ~\cite{HMNcikk} for this model at different parameters.
By repeating the simulations at different sizes: $l=5,6$ the
distribution curves do not change within error margin, and this size
invariance implies the presence of real GP-s.

The seminal experiments by Beggs and Plenz \cite{BP03} reported
neuronal avalanches with size ($s$) dependence, defined as either the 
number of electrodes with 
supra-threshold activity or the sum of the potentials, according to a power
law, $p(s) \propto s^{-1.5}$. For the duration distribution of such events
$P(t) \propto t^{-2}$, PL tails were observed. These exponents are in agreement
with the mean-field (MF) exponents of the Directed Percolation (DP) criticality:
$\tau = 3/2$, $\tau_t = 2$ see \cite{rmp}.
Mean-field exponents are expected to occur if the fluctuation effects are
weak, when the system dimension is above the upper critical dimension
$d_c$. 

On the other hand, Palva et al.~\cite{brainexp} have found that 
source-reconstructed M/EEG data exhibit robust power-law, long-range
time correlations and scale-invariant avalanches with a broad range of 
exponents: $1 < \tau < 1.6$ and $1.5 < \tau_t < 2.4$. These broad
range exponent results have also been found in a recent cortical 
electrode experiment study on rodents~\cite{CC18}.
An obvious explanation for this wide spread of critical exponents
can be heterogeneity, which in the GP causes non-universal dynamical
exponents \cite{Moretti2013,HMNcikk,CCdyncikk,CC-tdepcikk}.


\subsection{Inhibitory model}


Although inhibitory links are not expected at long range links of the 
brain~\cite{LatNir2004}, we believe that our synthetic model may describe 
smaller cortical scales as well. 
Besides, inhibitory mechanisms can occur in other phenomena with 
threshold dynamics. In case of power-grids, for example, feedback 
is applied to prevent catastrophic blackout avalanches, or in models of
social/epidemic contagion, nodes with inhibitory properties 
may also exist. For simplicity we modeled the inhibitions by the 
introduction of links with negative weight contribution ($w_{i,j}$)
in the threshold comparison rule given by Eq.~\ref{thresh},
although we think our results are easily transferred to the case of
inhibitory nodes. As in~\cite{CCdyncikk,CC-tdepcikk}, we randomly flipped 
the sign of $20\%$  of the edges after the generation
of the network.

\begin{figure}[h]
\includegraphics[height=6.5cm]{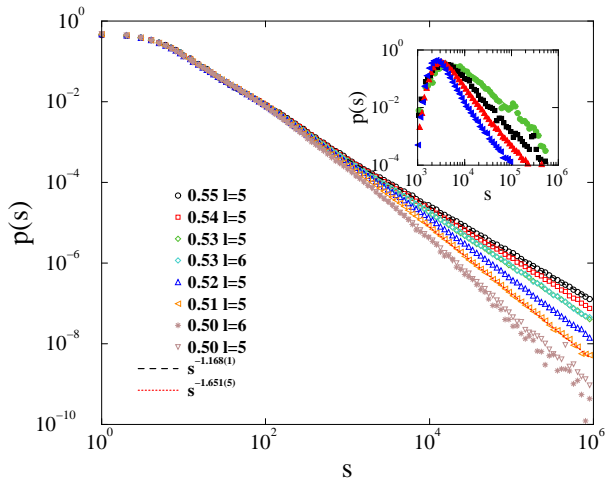}
\caption{Avalanche size distributions at different $\lambda$ branching rates, 
denoted by the symbols, in the presence of inhibitory links in HMN2d with 
$l=5,6$ levels. From top to bottom curves: $\lambda= $0.55, 0.54, 0.53 
($l=5$ green and $l=6$ cyan), 0.52, 0.51, 0.50 ($l=5$ triangle and 
$l=6$ diamond). 
Dashed lines show power-law fits for the tails of $\lambda=0.5, 0.6$ cases,
i.e. for $t > 1000$.
Inset: overlapping avalanches case for half filled initial condition at: 
$\lambda=0.51, 0.515, 0.52, 0.525$ (bottom to top symbols).}
\label{elo-hmni}
\end{figure}

The same analysis resulted in similar behavior as for the excitatory case.
One can see $p(s)$ distributions with non-universal PL-s ranging from
$\lambda=0.5$ with $\tau=1.651(1)$ to $\lambda=0.55$ with 
$\tau=1.168(1)$ (Fig.~\ref{elo-hmni}).
Finite size dependence is not visible by changing the size from $N=4096$ 
to $N=16384$. 

Usually it is believed that overlapping avalanches distort the
scaling behavior. From the point of view of statistical physics, this would
contradict universal asymptotic scaling behavior. Indeed we can see the same
cumulative $p(s)$ distribution tails even in case of starting the system
from half filled active state as shown on the inset of Fig.~\ref{elo-hmni}.
The only difference is that the tails are shifted to larger $s$ values 
following an initial growth, which might not be observed in case
of short time measurements.

\begin{figure}[h]
\includegraphics[height=6.5cm]{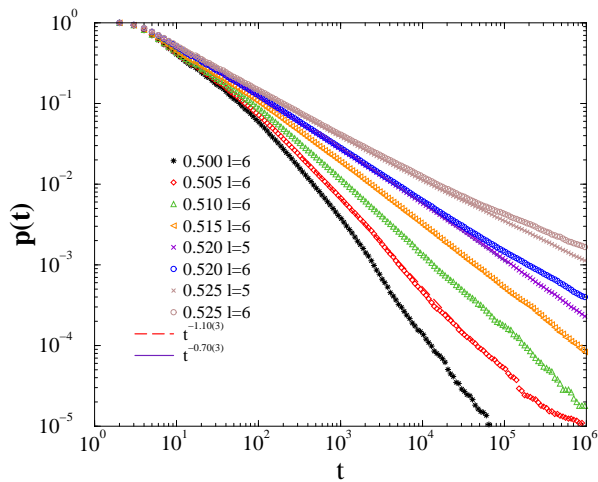}
\caption{\label{Phier21}Survival probability of the activity at different
branching rates and $\nu=1-\lambda$ for the $K=2$ threshold model with
levels: $l=5,6$ for the case with $20\%$ of inhibitory links. 
From bottom to top symbols: $\lambda=$ 0.5, 0.505, 0.510, 0.515, 0.520 
($l=5$ purple cross and $l=6$ blue circle), 0.525 ($l=5$ brown cross 
and $l=6$ brown circle).  }
\end{figure}

The $p(t)$ decays show GP behavior from $\lambda=0.505$ with $\delta=1.10(3)$  
to $\lambda=0.52$ with $\delta=0.70(3)$ (Fig.~\ref{Phier21}).
These values do not correspond to the ends of the GP, we did not aim 
to determine them precisely as the exponents are non-universal.
Furthermore, as we will show in Sect.~\ref{sss:sect} the determination
of the upper limit of the GP, corresponding to the critical decay is
almost impossible by numerical simulations.

Again the $\tau$ and the $\tau_t=1+\delta$ values lie within the
range obtained by experiments.


\subsection{Inhibitory-refractory model}

\begin{figure}[h]
\includegraphics[height=6.5cm]{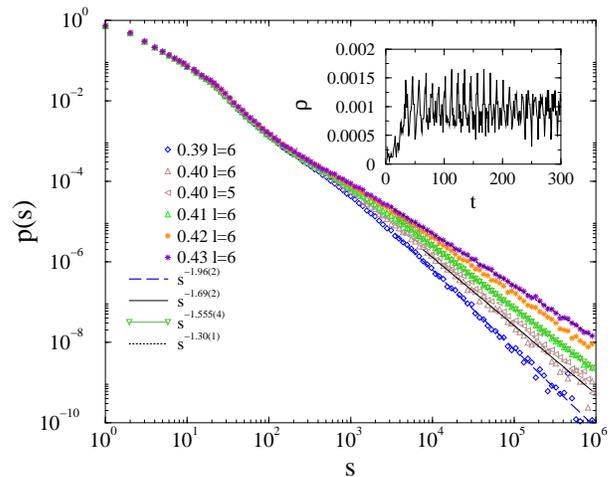}
\caption{Avalanche size distributions at different $\lambda$ branching rates, 
denoted by the symbols, in case of the refractory model,
in the presence of inhibitory links in HMN2d-s with $l=5,6$ levels. 
From bottom to top symbols: $\lambda=$ 0.39, 0.40 ($l=5$ left triangle and 
$l=6$ up triangle), 0.41, 0.42, 0.43. 
Dashed lines are PL fits for the tails of $\lambda=0.5, 0.6$ cases 
for $t>1000$. The inset shows the oscillatory behavior of $\rho(t)$ 
of a single run for $\Delta t = 10$.}
\label{elo-hmnIR}
\end{figure}

Finally, we extended the inhibitory case study with the possibility of
refractory states. Refractorieness means that, following an activation, 
nodes cannot fall back immediately to inactive state on the next update, 
instead they stay for time $\Delta t$ in a refractory state. 
Thus they cannot be reactivated by the neighbors they excited.
This refractoriness is generic in excitable systems and has been used in 
most of the neural studies~\cite{PhysRevLett.110.178101,Afshin,PhysRevE.100.052138}.
One of the consequences of refractoriness is to induce oscillatory 
dynamics if $\Delta t$ is large enough and the spreading properties 
resemble to annular growth, corresponding to Dynamical Isotropic 
Percolation (DIP)~\cite{odorbook}.
However, real DIP occurs if re-activation is not possible, i.e.
in the limit $\Delta t \to \infty$, and in high dimension the avalanche 
scaling exponents of DIP are the same as those of DP~\cite{Munoz99,odorbook}. 
Thus analytic studies or simulations in high dimensions do not show 
differences. In the extensive GP simulations we used $\Delta t=1$, but on the 
inset of Fig.~\ref{elo-hmnIR} we show oscillatory activity
behavior of a single run for $\Delta t = 10$, $\lambda=0.8$ and $l=6$.

In \cite{CC-tdepcikk} the GP behavior of inhibitory-refractory threshold
model was investigated on a large human connectome graph numerically.
Non-universal $p(s)$ decays were reported with $1.4 < \tau < 1.91$.
Here we can see this in the range $\lambda=0.39$ with $\tau=1.96(2)$,
to $\lambda=0.43$ with $\tau=1.39(1)$. 

\begin{figure}[h]
\includegraphics[height=6.5cm]{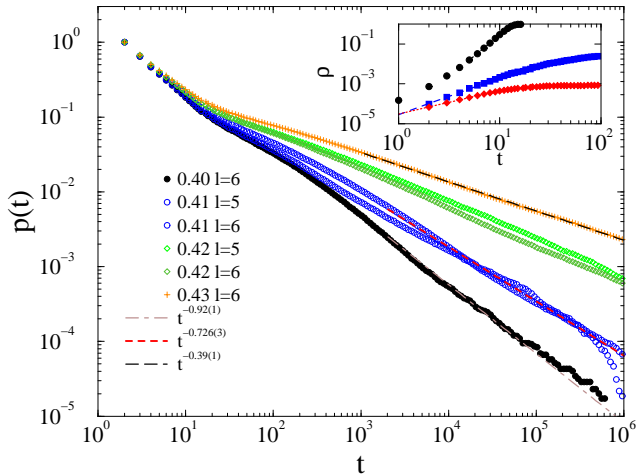}
\caption{\label{Pir3l21}Survival probability of the activity at different
branching rates $\lambda$ for the levels: $l=5,6$, in case of the inhibitory-refractory  
model. From bottom to top symbols: $\lambda=$0.40, 0.41 ($l=5$ and $l=6$), 
0.42 ($l=5$ light green and $l=6$ dark green), 0.43. Dashed lines show
PL fits for $t > 1000$ for the $\lambda=$ 0.4, 0.41, 0.43 cases.
Inset:
$\rho(t)$ at $\lambda=1$, $l=7$ averaged over $10^5$ realizations. Blue boxes:
excitatory, red diamonds: inhibitory. Black bullets: BFS $\rho(r)$
results. Dashed lines are PL fits for the initial regions: $1 \le t < 10$)
resulting in effective dimensions: $d_{eff}=1.84(3)$ (excitatory),
$d_{eff}=1.19(1)$ (inhibitory), $d=4.18(5)$ (graph dimension estimated
for $5 < r < 10$).}
\end{figure}

The avalanche survival probabilities (see Fig.~\ref{Pir3l21}),
exhibit PL decay from $\lambda=0.40$ with $\delta=0.92(1)$, 
to $\lambda=0.43$ with $\delta=0.39(1)$,
so the duration exponent varies continuously: 
$1.39(1) \le \tau_t \le 1.92(1)$.
Note that for similar models in ~\cite{Afshin,PhysRevE.100.052138} 
complex phase diagrams and non-universal PL-s have also been found and 
the possibility of GP has been pointed out.


\section{Steady state simulations}\label{sss:sect}

\begin{figure}[h]
\includegraphics[height=6.5cm]{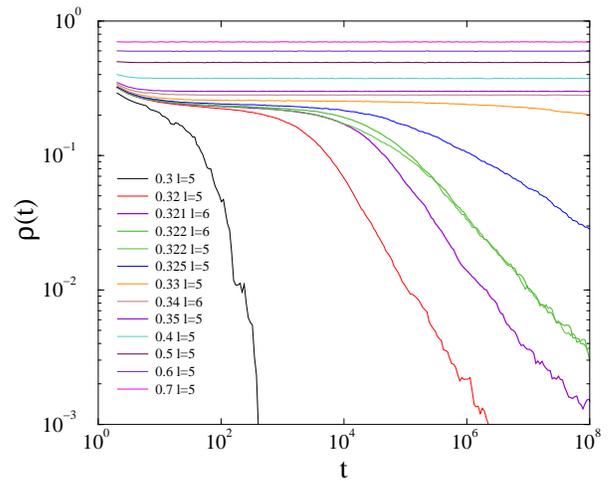}
\caption{Evolution of $\rho(t)$ for different $\lambda$-s in case of
starting from fully active state in the excitatory model with levels: $l=5,6$. 
From bottom to top symbols: $\lambda=$ 0.30, 0.32, 0.321 ($l=6$), 0.322, 0.322 ($l=6$), 
0.325, 0.33, 0.34 (l=6), 0.35, 0.4, 0.5, 0.6, 0.7.}
\label{rho21_N=4096_k=2}
\end{figure}
\begin{figure}[h]
\includegraphics[height=6.5cm]{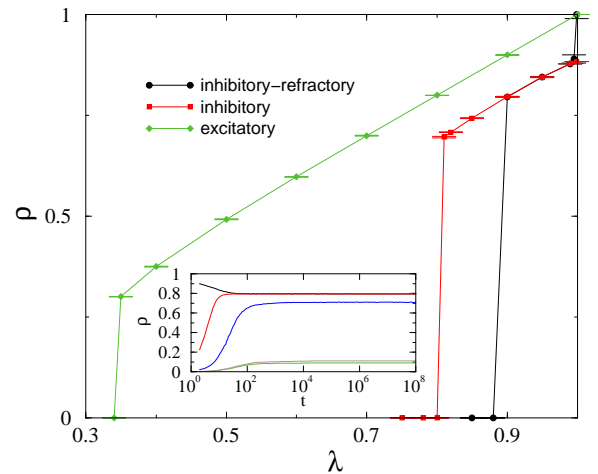}
\caption{Steady state behavior for the excitatory, inhibitory and
refractory-inhibitory cases. Inset: evolution of $\rho$ in an
inhibitory HMN2d with $N=4096$ for different initial activity densities:
$\rho(0)=$ 0.0005, 0.001, 0.01, 0.1, 1 (bottom to top curves).}
\label{betafig}
\end{figure}

In order to determine the steady state behavior we first performed long runs,
up to $T=10^8$ MCs, by starting the system from fully active state or
from randomly half filled activity: $\rho(0)=0.5$.
Fig.~\ref{rho21_N=4096_k=2} shows the results for the excitatory model. 
At $\lambda = 0.3$ the activity density falls exponentially fast to zero.
We can see non-universal PL tails for $0.32 \le \lambda < 0.33$,
in agreement with the seed simulations. At $\lambda=0.33$ the density does
not saturate to a constant value. Examining it on log.-lin. scale and
performing average over thousands of independent samples it turns out
that even the $\lambda=0.34$ curve decays very slowly. 
Only for $\lambda \ge 0.35$ we can see saturation, corresponding to
active steady state, thus, we estimate $\lambda_c = 0.345(5)$. 
We plotted the steady state saturation values on Fig.~\ref{betafig}. 

\begin{figure}[h]
\includegraphics[height=6.5cm]{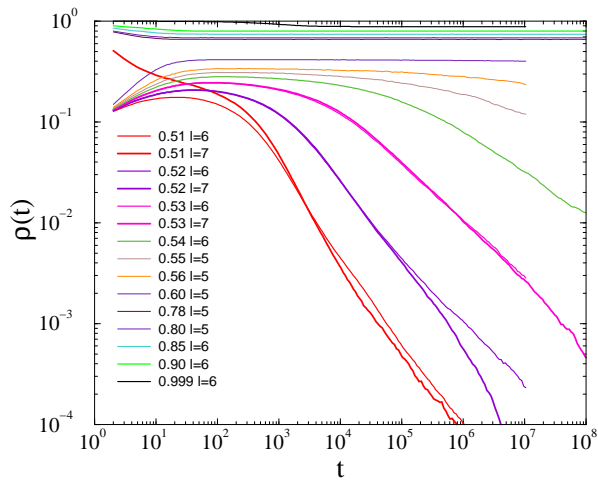}
\caption{Evolution of $\rho(t)$ for different $\lambda$-s, shown by
legends in case of starting from active states in the inhibitory model.
Thick, normal, thin lines correspond to $l=7$, $l=6$, $l=5$, accordingly.
From bottom to top curves: $\lambda=$ 0.51, 0.52, 0.53, 0.54, 0.55, 0.56, 0.6,
0.78, 0.8, 0.85, 0.9, 0.999. The graph shows results with initial condition
$\rho(0)=0.5$ for $\lambda \le 0.6$, except for $\lambda= 0.51$, $l=7$.
In all the other cases $\rho(0)=1$ is applied. }
\label{rhosihier2-GP}
\end{figure}

The same analysis has been done for the inhibitory and refractory-inhibitory 
cases and one can observe the shift of $\lambda_c$ to higher values as the
consequence of the model modifications. We show the results for
the inhibitory network on Fig.~\ref{rhosihier2-GP}.
Again, slow activity decays were observed, ending up with visible PL tails for 
$0.51 \le\lambda\le 0.54$, while saturation starts from $\lambda_c\ge0.80(1)$.
The saturation value is $\rho_c=0.685(1)$, so the discontinuity is large.
In the region $0.54 < \lambda < 0.80$ the curves do not saturate
up to $T=10^8$ MCs, neither reach a scaling region. They belong to the
inactive phase, but it is very difficult to distinguish them from 
other (logarithmic) decay forms.

We can see large jumps at the transition points in all cases, 
suggesting a discontinuous transition above the GP. It is very hard to
locate the exact location of the transition points as stability 
disappears very slowly. This suggests that at the critical point 
logarithmic decay occurs like in case of the disordered 
DP~\cite{Vojta2006b}.

We have also tried to start from other initial conditions than
the full and single seed ones. As the inset of Fig.~\ref{betafig} 
shows we can see different saturation values for $\rho(0)<0.1$, that means
we have multi-stability in case of low initial densities.
This is the consequence of the fact that for low $\rho(0)$-s 
only parts of the graph can be activated. Even though the networks 
are simple connected and the lowest in-degree is $k_i^{in}=5$, 
not all nodes have $2$ incoming links from the same neighbor, which
is necessary for the activation.
These nodes cannot be activated by a neighbor if they are on the 
"border" of an active territory, thus the graphs are practically 
fragmented from the activity point of view.
This provides a mechanism for the emergence of GP even in high
dimensions, without breaking conjecture provided
in~\cite{Munoz2010}, according to which GPs and similar RR 
effects do not exist in networks with an infinite topological dimension.

Furthermore, we can see the emergence of discontinuous transition with
multi-stable states, which can be considered bi-stable, for initial
excitation with node fraction $\rho(0) > 0.1$ converging to an "up" 
activity value, or by activation of single nodes, 
converging to a "down" value.


\section{Conclusions}


In conclusion we provided numerical evidence that strong heterogeneity
effects in networks, coming from the modular structure can result in
GP even if the topological dimension is high, where mean-field scaling
would be expected. This is the consequence of fragmented activity 
propagation caused by the modular topology and the threshold.
We can define effective dimensions of the these graphs
by running seed simulations with $\lambda=1$, $\nu=0$ and
measuring $\rho(t) \sim t^{d_{eff}}$. For this compact growth
$\rho(t) \sim N(r)$, so $d_{eff}$ provides an estimate for the
dimensionality, similarly to the BFS algorithm.
This is reminiscent of similar methods, for instance computing 
the spectral dimension of a network from random walk simulations
~\cite{PhysRevLett.76.1091,Burioni_2000,Burioni_2005,PhysRevE.99.022307}.
While the topological dimension is a purely structural measure, 
$d_{eff}$, as well as the spectral dimension are observables
of processes operating on networks, providing insights to 
dynamical signatures of localization, slowing down and 
dynamical fragmentation. However, it has recently been shown 
that in models of HMN-s the spectral dimension is not 
defined~\cite{PhysRevResearch.2.043291}, thus our $d_{eff}$
can be a candidate to clarify relation of structure and
slow dynamics. 
The inset of Fig.~\ref{Pir3l21} shows that an initial scaling
can be fitted for the excitatory case with: $\rho(t)\sim t^{1.84(3)}$,
while for inhibitory: $\rho(t)\sim t^{1.19(1)}$. Thus these effective,
activity dimensions are less than $d_c$, much smaller than the topological 
dimension obtained by the BFS, which is also shown on
the graph as the function of $r$.

Furthermore, the threshold type models allow for the possibility to observe
hybrid phase transitions, where order parameter discontinuity and 
multi-stability can co-exist with dynamical scaling in a GP, thus
they can model brain criticality as well as up/down states.
External activation can then push the model among the multi-stable 
states if it is poised near the transition point.

The investigated $K=2$ discrete threshold model results can obviously 
be extended for higher $K$ values and we expect to find similar behavior
in continuous, integrate and fire type models on modular networks.
Conversely, by duplicating the links we get back effectively the contact 
process~\cite{Harris74} without RR effects and GP. 
For neural systems our results
imply that the functional and structural connectivity can be different.
The effects of inhibition and refractive states have also been studied
and emergence of oscillatory states have been shown.
Our model results are applicable to a wide range of phenomena, like
power-grids, crack and fracture dynamics and contagion.

\section*{Acknowledgments}

We thank R\'obert Juh\'asz for useful comments and discussions,
Miguel Mu\~noz and Dante Chialvo for their valuable feedbacks.
Support from the Hungarian National Research, Development and Innovation
Office NKFIH (K128989) is acknowledged. We thank access to the
Hungarian National Supercomputer Network.

\bibliography{main}

\begin{thebibliography}{73}
\expandafter\ifx\csname natexlab\endcsname\relax\def\natexlab#1{#1}\fi
\expandafter\ifx\csname bibnamefont\endcsname\relax
  \def\bibnamefont#1{#1}\fi
\expandafter\ifx\csname bibfnamefont\endcsname\relax
  \def\bibfnamefont#1{#1}\fi
\expandafter\ifx\csname citenamefont\endcsname\relax
  \def\citenamefont#1{#1}\fi
\expandafter\ifx\csname url\endcsname\relax
  \def\url#1{\texttt{#1}}\fi
\expandafter\ifx\csname urlprefix\endcsname\relax\def\urlprefix{URL }\fi
\providecommand{\bibinfo}[2]{#2}
\providecommand{\eprint}[2][]{\url{#2}}

\bibitem[{\citenamefont{Marro and Dickman}(2005)}]{marro2005}
\bibinfo{author}{\bibfnamefont{J.}~\bibnamefont{Marro}} \bibnamefont{and}
  \bibinfo{author}{\bibfnamefont{R.}~\bibnamefont{Dickman}},
  \emph{\bibinfo{title}{Nonequilibrium Phase Transitions in Lattice Models}},
  Al{\'e}a-Saclay (\bibinfo{publisher}{Cambridge University Press},
  \bibinfo{year}{2005}), ISBN \bibinfo{isbn}{9780521019460}.

\bibitem[{\citenamefont{Henkel et~al.}(2008)\citenamefont{Henkel, Hinrichsen,
  and L\"ubeck}}]{HHL}
\bibinfo{author}{\bibfnamefont{M.}~\bibnamefont{Henkel}},
  \bibinfo{author}{\bibfnamefont{H.}~\bibnamefont{Hinrichsen}},
  \bibnamefont{and} \bibinfo{author}{\bibfnamefont{S.}~\bibnamefont{L\"ubeck}},
  \emph{\bibinfo{title}{Non-equilibrium phase transition: Absorbing Phase
  Transitions}} (\bibinfo{publisher}{Springer Verlag},
  \bibinfo{address}{Netherlands}, \bibinfo{year}{2008}).

\bibitem[{\citenamefont{\'Odor}(2004)}]{rmp}
\bibinfo{author}{\bibfnamefont{G.}~\bibnamefont{\'Odor}},
  \bibinfo{journal}{Reviews of Modern Physics} \textbf{\bibinfo{volume}{76}},
  \bibinfo{pages}{663} (\bibinfo{year}{2004}).

\bibitem[{\citenamefont{{\'O}dor}(2008)}]{odorbook}
\bibinfo{author}{\bibfnamefont{G.}~\bibnamefont{{\'O}dor}},
  \emph{\bibinfo{title}{Universality in nonequilibrium lattice systems:
  Theoretical foundations}} (\bibinfo{publisher}{World Scientific},
  \bibinfo{year}{2008}).

\bibitem[{\citenamefont{Cardy}(1981)}]{Cardy_1981}
\bibinfo{author}{\bibfnamefont{J.~L.} \bibnamefont{Cardy}},
  \bibinfo{journal}{Journal of Physics A: Mathematical and General}
  \textbf{\bibinfo{volume}{14}}, \bibinfo{pages}{1407} (\bibinfo{year}{1981}).

\bibitem[{\citenamefont{Janssen et~al.}(2004)\citenamefont{Janssen, M\"uller,
  and Stenull}}]{PhysRevE.70.026114}
\bibinfo{author}{\bibfnamefont{H.-K.} \bibnamefont{Janssen}},
  \bibinfo{author}{\bibfnamefont{M.}~\bibnamefont{M\"uller}}, \bibnamefont{and}
  \bibinfo{author}{\bibfnamefont{O.}~\bibnamefont{Stenull}},
  \bibinfo{journal}{Physical Review E} \textbf{\bibinfo{volume}{70}},
  \bibinfo{pages}{026114} (\bibinfo{year}{2004}).

\bibitem[{\citenamefont{Chan et~al.}(2015)\citenamefont{Chan, Ghanbarnejad, and
  Grassberger}}]{Gras-av}
\bibinfo{author}{\bibfnamefont{W.}~\bibnamefont{Chan}},
  \bibinfo{author}{\bibfnamefont{F.}~\bibnamefont{Ghanbarnejad}},
  \bibnamefont{and}
  \bibinfo{author}{\bibfnamefont{P.}~\bibnamefont{Grassberger}},
  \bibinfo{journal}{Nature Physics} \textbf{\bibinfo{volume}{11}},
  \bibinfo{pages}{936–} (\bibinfo{year}{2015}).

\bibitem[{\citenamefont{Dorogovtsev et~al.}(2006)\citenamefont{Dorogovtsev,
  Goltsev, and Mendes}}]{PhysRevLett.96.040601}
\bibinfo{author}{\bibfnamefont{S.~N.} \bibnamefont{Dorogovtsev}},
  \bibinfo{author}{\bibfnamefont{A.~V.} \bibnamefont{Goltsev}},
  \bibnamefont{and} \bibinfo{author}{\bibfnamefont{J.~F.~F.}
  \bibnamefont{Mendes}}, \bibinfo{journal}{Physical Review Letters}
  \textbf{\bibinfo{volume}{96}}, \bibinfo{pages}{040601}
  (\bibinfo{year}{2006}).

\bibitem[{\citenamefont{Lee et~al.}(2016)\citenamefont{Lee, Choi, Stippinger,
  Kert\'esz, and Kahng}}]{PhysRevE.93.042109}
\bibinfo{author}{\bibfnamefont{D.}~\bibnamefont{Lee}},
  \bibinfo{author}{\bibfnamefont{S.}~\bibnamefont{Choi}},
  \bibinfo{author}{\bibfnamefont{M.}~\bibnamefont{Stippinger}},
  \bibinfo{author}{\bibfnamefont{J.}~\bibnamefont{Kert\'esz}},
  \bibnamefont{and} \bibinfo{author}{\bibfnamefont{B.}~\bibnamefont{Kahng}},
  \bibinfo{journal}{Physical Review E} \textbf{\bibinfo{volume}{93}},
  \bibinfo{pages}{042109} (\bibinfo{year}{2016}).

\bibitem[{\citenamefont{Baxter et~al.}(2014)\citenamefont{Baxter, Dorogovtsev,
  Mendes, and Cellai}}]{PhysRevE.89.042801}
\bibinfo{author}{\bibfnamefont{G.~J.} \bibnamefont{Baxter}},
  \bibinfo{author}{\bibfnamefont{S.~N.} \bibnamefont{Dorogovtsev}},
  \bibinfo{author}{\bibfnamefont{J.~F.~F.} \bibnamefont{Mendes}},
  \bibnamefont{and} \bibinfo{author}{\bibfnamefont{D.}~\bibnamefont{Cellai}},
  \bibinfo{journal}{Physical Review E} \textbf{\bibinfo{volume}{89}},
  \bibinfo{pages}{042801} (\bibinfo{year}{2014}).

\bibitem[{\citenamefont{Bar and Mukamel}(2014)}]{PhysRevLett.112.015701}
\bibinfo{author}{\bibfnamefont{A.}~\bibnamefont{Bar}} \bibnamefont{and}
  \bibinfo{author}{\bibfnamefont{D.}~\bibnamefont{Mukamel}},
  \bibinfo{journal}{Physical Review Letters} \textbf{\bibinfo{volume}{112}},
  \bibinfo{pages}{015701} (\bibinfo{year}{2014}).

\bibitem[{\citenamefont{Juh\'asz and Igl\'oi}(2017)}]{PhysRevE.95.022109}
\bibinfo{author}{\bibfnamefont{R.}~\bibnamefont{Juh\'asz}} \bibnamefont{and}
  \bibinfo{author}{\bibfnamefont{F.}~\bibnamefont{Igl\'oi}},
  \bibinfo{journal}{Physical Review E} \textbf{\bibinfo{volume}{95}},
  \bibinfo{pages}{022109} (\bibinfo{year}{2017}).

\bibitem[{\citenamefont{Grassberger}(2017)}]{PhysRevE.95.010102}
\bibinfo{author}{\bibfnamefont{P.}~\bibnamefont{Grassberger}},
  \bibinfo{journal}{Physical Review E} \textbf{\bibinfo{volume}{95}},
  \bibinfo{pages}{010102(R)} (\bibinfo{year}{2017}).

\bibitem[{\citenamefont{Juh\'asz and Igl\'oi}(2018)}]{PhysRevE.97.012111}
\bibinfo{author}{\bibfnamefont{R.}~\bibnamefont{Juh\'asz}} \bibnamefont{and}
  \bibinfo{author}{\bibfnamefont{F.}~\bibnamefont{Igl\'oi}},
  \bibinfo{journal}{Physical Review E} \textbf{\bibinfo{volume}{97}},
  \bibinfo{pages}{012111} (\bibinfo{year}{2018}).

\bibitem[{\citenamefont{Paz\'o}(2005)}]{PhysRevE.72.046211}
\bibinfo{author}{\bibfnamefont{D.}~\bibnamefont{Paz\'o}},
  \bibinfo{journal}{Physical Review E} \textbf{\bibinfo{volume}{72}},
  \bibinfo{pages}{046211} (\bibinfo{year}{2005}).

\bibitem[{\citenamefont{G\'omez-Garde\~nes
  et~al.}(2011)\citenamefont{G\'omez-Garde\~nes, G\'omez, Arenas, and
  Moreno}}]{PhysRevLett.106.128701}
\bibinfo{author}{\bibfnamefont{J.}~\bibnamefont{G\'omez-Garde\~nes}},
  \bibinfo{author}{\bibfnamefont{S.}~\bibnamefont{G\'omez}},
  \bibinfo{author}{\bibfnamefont{A.}~\bibnamefont{Arenas}}, \bibnamefont{and}
  \bibinfo{author}{\bibfnamefont{Y.}~\bibnamefont{Moreno}},
  \bibinfo{journal}{Physical Review Letters} \textbf{\bibinfo{volume}{106}},
  \bibinfo{pages}{128701} (\bibinfo{year}{2011}).

\bibitem[{\citenamefont{Coutinho et~al.}(2013)\citenamefont{Coutinho, Goltsev,
  Dorogovtsev, and Mendes}}]{PhysRevE.87.032106}
\bibinfo{author}{\bibfnamefont{B.~C.} \bibnamefont{Coutinho}},
  \bibinfo{author}{\bibfnamefont{A.~V.} \bibnamefont{Goltsev}},
  \bibinfo{author}{\bibfnamefont{S.~N.} \bibnamefont{Dorogovtsev}},
  \bibnamefont{and} \bibinfo{author}{\bibfnamefont{J.~F.~F.}
  \bibnamefont{Mendes}}, \bibinfo{journal}{Physical Review E}
  \textbf{\bibinfo{volume}{87}}, \bibinfo{pages}{032106}
  (\bibinfo{year}{2013}).

\bibitem[{\citenamefont{Haimovici et~al.}(2013)\citenamefont{Haimovici,
  Tagliazucchi, Balenzuela, and Chialvo}}]{PhysRevLett.110.178101}
\bibinfo{author}{\bibfnamefont{A.}~\bibnamefont{Haimovici}},
  \bibinfo{author}{\bibfnamefont{E.}~\bibnamefont{Tagliazucchi}},
  \bibinfo{author}{\bibfnamefont{P.}~\bibnamefont{Balenzuela}},
  \bibnamefont{and} \bibinfo{author}{\bibfnamefont{D.~R.}
  \bibnamefont{Chialvo}}, \bibinfo{journal}{Physical Review Letters}
  \textbf{\bibinfo{volume}{110}}, \bibinfo{pages}{178101}
  (\bibinfo{year}{2013}).

\bibitem[{\citenamefont{Johnson~S. and Marro}(2013)}]{Johnson}
\bibinfo{author}{\bibfnamefont{J.~J.} \bibnamefont{Johnson~S.},
  \bibfnamefont{Torres}} \bibnamefont{and}
  \bibinfo{author}{\bibfnamefont{J.}~\bibnamefont{Marro}},
  \bibinfo{journal}{PLoS ONE} \textbf{\bibinfo{volume}{8}},
  \bibinfo{pages}{e50276} (\bibinfo{year}{2013}).

\bibitem[{\citenamefont{Bak et~al.}(1987)\citenamefont{Bak, Tang, and
  Wiesenfeld}}]{SOC}
\bibinfo{author}{\bibfnamefont{P.}~\bibnamefont{Bak}},
  \bibinfo{author}{\bibfnamefont{C.}~\bibnamefont{Tang}}, \bibnamefont{and}
  \bibinfo{author}{\bibfnamefont{K.}~\bibnamefont{Wiesenfeld}},
  \bibinfo{journal}{Physical Review Letters} \textbf{\bibinfo{volume}{59}},
  \bibinfo{pages}{381} (\bibinfo{year}{1987}).

\bibitem[{\citenamefont{Chialvo}(2010)}]{Chialvo2010}
\bibinfo{author}{\bibfnamefont{D.~R.} \bibnamefont{Chialvo}},
  \bibinfo{journal}{Nature Physics} \textbf{\bibinfo{volume}{6}},
  \bibinfo{pages}{744} (\bibinfo{year}{2010}).

\bibitem[{\citenamefont{Vojta}(2006)}]{Vojta2006b}
\bibinfo{author}{\bibfnamefont{T.}~\bibnamefont{Vojta}},
  \bibinfo{journal}{Journal of Physics A: Mathematical and General}
  \textbf{\bibinfo{volume}{39}}, \bibinfo{pages}{R143} (\bibinfo{year}{2006}).

\bibitem[{\citenamefont{Griffiths}(1969)}]{Griffiths}
\bibinfo{author}{\bibfnamefont{R.~B.} \bibnamefont{Griffiths}},
  \bibinfo{journal}{Physical Review Letters} \textbf{\bibinfo{volume}{23}},
  \bibinfo{pages}{17} (\bibinfo{year}{1969}), ISSN \bibinfo{issn}{0031-9007}.

\bibitem[{\citenamefont{{\'O}dor}(2014)}]{burstcikk}
\bibinfo{author}{\bibfnamefont{G.}~\bibnamefont{{\'O}dor}},
  \bibinfo{journal}{Physical Review E} \textbf{\bibinfo{volume}{89}},
  \bibinfo{pages}{042102} (\bibinfo{year}{2014}).

\bibitem[{\citenamefont{Karsai et~al.}(2018)\citenamefont{Karsai, Jo, and
  Kaski}}]{Karsai_2018}
\bibinfo{author}{\bibfnamefont{M.}~\bibnamefont{Karsai}},
  \bibinfo{author}{\bibfnamefont{H.-H.} \bibnamefont{Jo}}, \bibnamefont{and}
  \bibinfo{author}{\bibfnamefont{K.}~\bibnamefont{Kaski}},
  \bibinfo{journal}{SpringerBriefs in Complexity}  (\bibinfo{year}{2018}), ISSN
  \bibinfo{issn}{2191-5334}.

\bibitem[{\citenamefont{Cota et~al.}(2016)\citenamefont{Cota, Ferreira, and
  {\'{O}}dor}}]{Cota2016}
\bibinfo{author}{\bibfnamefont{W.}~\bibnamefont{Cota}},
  \bibinfo{author}{\bibfnamefont{S.~C.} \bibnamefont{Ferreira}},
  \bibnamefont{and}
  \bibinfo{author}{\bibfnamefont{G.}~\bibnamefont{{\'{O}}dor}},
  \bibinfo{journal}{Physical Review E} \textbf{\bibinfo{volume}{93}},
  \bibinfo{pages}{032322} (\bibinfo{year}{2016}).

\bibitem[{\citenamefont{Imry and Ma}(1975)}]{IM}
\bibinfo{author}{\bibfnamefont{Y.}~\bibnamefont{Imry}} \bibnamefont{and}
  \bibinfo{author}{\bibfnamefont{S.-k.} \bibnamefont{Ma}},
  \bibinfo{journal}{Physical Review Letters} \textbf{\bibinfo{volume}{35}},
  \bibinfo{pages}{1399} (\bibinfo{year}{1975}).

\bibitem[{\citenamefont{Aizenman and Wehr}(1989)}]{Aw}
\bibinfo{author}{\bibfnamefont{M.}~\bibnamefont{Aizenman}} \bibnamefont{and}
  \bibinfo{author}{\bibfnamefont{J.}~\bibnamefont{Wehr}},
  \bibinfo{journal}{Physical Review Letters} \textbf{\bibinfo{volume}{62}},
  \bibinfo{pages}{2503} (\bibinfo{year}{1989}).

\bibitem[{\citenamefont{Villa~Mart\'{\i}n
  et~al.}(2014)\citenamefont{Villa~Mart\'{\i}n, Bonachela, and
  Mu\~noz}}]{round}
\bibinfo{author}{\bibfnamefont{P.}~\bibnamefont{Villa~Mart\'{\i}n}},
  \bibinfo{author}{\bibfnamefont{J.~A.} \bibnamefont{Bonachela}},
  \bibnamefont{and} \bibinfo{author}{\bibfnamefont{M.~A.}
  \bibnamefont{Mu\~noz}}, \bibinfo{journal}{Physical Review E}
  \textbf{\bibinfo{volume}{89}}, \bibinfo{pages}{012145}
  (\bibinfo{year}{2014}).

\bibitem[{\citenamefont{Villa~Mart\'{\i}n
  et~al.}(2015)\citenamefont{Villa~Mart\'{\i}n, Moretti, and Mu\~noz}}]{round2}
\bibinfo{author}{\bibfnamefont{P.}~\bibnamefont{Villa~Mart\'{\i}n}},
  \bibinfo{author}{\bibfnamefont{M.}~\bibnamefont{Moretti}}, \bibnamefont{and}
  \bibinfo{author}{\bibfnamefont{M.~A.} \bibnamefont{Mu\~noz}},
  \bibinfo{journal}{Journal of Statistical Mechanics} p.
  \bibinfo{pages}{P01003} (\bibinfo{year}{2015}).

\bibitem[{\citenamefont{Beggs and Plenz}(2003)}]{BP03}
\bibinfo{author}{\bibfnamefont{J.}~\bibnamefont{Beggs}} \bibnamefont{and}
  \bibinfo{author}{\bibfnamefont{D.}~\bibnamefont{Plenz}},
  \bibinfo{journal}{Journal Neuroscience} \textbf{\bibinfo{volume}{23}},
  \bibinfo{pages}{11167} (\bibinfo{year}{2003}).

\bibitem[{\citenamefont{Tetzlaff et~al.}(2010)\citenamefont{Tetzlaff, Okujeni,
  Egert, W{\"o}rg{\"o}tter, and Butz}}]{T10}
\bibinfo{author}{\bibfnamefont{C.}~\bibnamefont{Tetzlaff}},
  \bibinfo{author}{\bibfnamefont{S.}~\bibnamefont{Okujeni}},
  \bibinfo{author}{\bibfnamefont{U.}~\bibnamefont{Egert}},
  \bibinfo{author}{\bibfnamefont{F.}~\bibnamefont{W{\"o}rg{\"o}tter}},
  \bibnamefont{and} \bibinfo{author}{\bibfnamefont{M.}~\bibnamefont{Butz}},
  \bibinfo{journal}{PLoS Computational Biology} \textbf{\bibinfo{volume}{6}},
  \bibinfo{pages}{e1001013} (\bibinfo{year}{2010}).

\bibitem[{\citenamefont{Hahn et~al.}(2010)\citenamefont{Hahn, Petermann,
  Havenith, Yu, Singer, Plenz, and Nikolić}}]{H10}
\bibinfo{author}{\bibfnamefont{G.}~\bibnamefont{Hahn}},
  \bibinfo{author}{\bibfnamefont{T.}~\bibnamefont{Petermann}},
  \bibinfo{author}{\bibfnamefont{M.~N.} \bibnamefont{Havenith}},
  \bibinfo{author}{\bibfnamefont{S.}~\bibnamefont{Yu}},
  \bibinfo{author}{\bibfnamefont{W.}~\bibnamefont{Singer}},
  \bibinfo{author}{\bibfnamefont{D.}~\bibnamefont{Plenz}}, \bibnamefont{and}
  \bibinfo{author}{\bibfnamefont{D.}~\bibnamefont{Nikolić}},
  \bibinfo{journal}{Journal of Neurophysiology} \textbf{\bibinfo{volume}{104}},
  \bibinfo{pages}{3312} (\bibinfo{year}{2010}), \bibinfo{note}{pMID: 20631221}.

\bibitem[{\citenamefont{Ribeiro et~al.}(2010)\citenamefont{Ribeiro, Copelli,
  Caixeta, Belchior, Chialvo, Nicolelis, and Riberio}}]{R10}
\bibinfo{author}{\bibfnamefont{T.}~\bibnamefont{Ribeiro}},
  \bibinfo{author}{\bibfnamefont{M.}~\bibnamefont{Copelli}},
  \bibinfo{author}{\bibfnamefont{F.}~\bibnamefont{Caixeta}},
  \bibinfo{author}{\bibfnamefont{H.}~\bibnamefont{Belchior}},
  \bibinfo{author}{\bibfnamefont{D.}~\bibnamefont{Chialvo}},
  \bibinfo{author}{\bibfnamefont{M.}~\bibnamefont{Nicolelis}},
  \bibnamefont{and} \bibinfo{author}{\bibfnamefont{S.}~\bibnamefont{Riberio}},
  \bibinfo{journal}{PLoS ONE} \textbf{\bibinfo{volume}{5}},
  \bibinfo{pages}{e14129} (\bibinfo{year}{2010}).

\bibitem[{\citenamefont{Ponce-Alvarez et~al.}(2018)\citenamefont{Ponce-Alvarez,
  Jouary, Privat, Deco, and Sumbre}}]{PONCEALVAREZ20181446}
\bibinfo{author}{\bibfnamefont{A.}~\bibnamefont{Ponce-Alvarez}},
  \bibinfo{author}{\bibfnamefont{A.}~\bibnamefont{Jouary}},
  \bibinfo{author}{\bibfnamefont{M.}~\bibnamefont{Privat}},
  \bibinfo{author}{\bibfnamefont{G.}~\bibnamefont{Deco}}, \bibnamefont{and}
  \bibinfo{author}{\bibfnamefont{G.}~\bibnamefont{Sumbre}},
  \bibinfo{journal}{Neuron} \textbf{\bibinfo{volume}{100}},
  \bibinfo{pages}{1446 } (\bibinfo{year}{2018}), ISSN
  \bibinfo{issn}{0896-6273}.

\bibitem[{\citenamefont{Mu\~noz}(2018)}]{MArep}
\bibinfo{author}{\bibfnamefont{M.~A.} \bibnamefont{Mu\~noz}},
  \bibinfo{journal}{Reviews of Modern Physics} \textbf{\bibinfo{volume}{90}},
  \bibinfo{pages}{031001} (\bibinfo{year}{2018}).

\bibitem[{\citenamefont{Durstewitz et~al.}(2000)\citenamefont{Durstewitz,
  Seamans, and Sejnowski}}]{bistab-exp}
\bibinfo{author}{\bibfnamefont{D.}~\bibnamefont{Durstewitz}},
  \bibinfo{author}{\bibfnamefont{J.~K.} \bibnamefont{Seamans}},
  \bibnamefont{and} \bibinfo{author}{\bibfnamefont{T.~J.}
  \bibnamefont{Sejnowski}}, \bibinfo{journal}{Nature Neuroscience}
  \textbf{\bibinfo{volume}{3}}, \bibinfo{pages}{1184} (\bibinfo{year}{2000}).

\bibitem[{\citenamefont{Kaltenbrunner et~al.}(2007)\citenamefont{Kaltenbrunner,
  G\'omez, and L\'opez}}]{doi:10.1162/neco.2007.19.11.3011}
\bibinfo{author}{\bibfnamefont{A.}~\bibnamefont{Kaltenbrunner}},
  \bibinfo{author}{\bibfnamefont{V.}~\bibnamefont{G\'omez}}, \bibnamefont{and}
  \bibinfo{author}{\bibfnamefont{V.}~\bibnamefont{L\'opez}},
  \bibinfo{journal}{Neural Computation} \textbf{\bibinfo{volume}{19}},
  \bibinfo{pages}{3011} (\bibinfo{year}{2007}).

\bibitem[{\citenamefont{Scarpetta et~al.}(2018)\citenamefont{Scarpetta,
  Apicella, Minati, and de~Candia}}]{PhysRevE.97.062305}
\bibinfo{author}{\bibfnamefont{S.}~\bibnamefont{Scarpetta}},
  \bibinfo{author}{\bibfnamefont{I.}~\bibnamefont{Apicella}},
  \bibinfo{author}{\bibfnamefont{L.}~\bibnamefont{Minati}}, \bibnamefont{and}
  \bibinfo{author}{\bibfnamefont{A.}~\bibnamefont{de~Candia}},
  \bibinfo{journal}{Physical Review E} \textbf{\bibinfo{volume}{97}},
  \bibinfo{pages}{062305} (\bibinfo{year}{2018}).

\bibitem[{\citenamefont{Buend{\'\i}a et~al.}(2020)\citenamefont{Buend{\'\i}a,
  Villegas, Burioni, and Mu{\~n}oz}}]{buendia2020hybridtype}
\bibinfo{author}{\bibfnamefont{V.}~\bibnamefont{Buend{\'\i}a}},
  \bibinfo{author}{\bibfnamefont{P.}~\bibnamefont{Villegas}},
  \bibinfo{author}{\bibfnamefont{R.}~\bibnamefont{Burioni}}, \bibnamefont{and}
  \bibinfo{author}{\bibfnamefont{M.~A.} \bibnamefont{Mu{\~n}oz}},
  \emph{\bibinfo{title}{Hybrid-type synchronization transitions: where marginal
  coherence, scale-free avalanches, and bistability live together}}
  (\bibinfo{year}{2020}), \eprint{arXiv:2011.03263}.

\bibitem[{\citenamefont{Kinouchi and Copelli}(2006)}]{KC}
\bibinfo{author}{\bibfnamefont{O.}~\bibnamefont{Kinouchi}} \bibnamefont{and}
  \bibinfo{author}{\bibfnamefont{M.}~\bibnamefont{Copelli}},
  \bibinfo{journal}{Nature Physics} \textbf{\bibinfo{volume}{2}},
  \bibinfo{pages}{348} (\bibinfo{year}{2006}).

\bibitem[{\citenamefont{Dobson et~al.}(2007)\citenamefont{Dobson, Carreras,
  Lynch, and Newman}}]{Car2}
\bibinfo{author}{\bibfnamefont{I.}~\bibnamefont{Dobson}},
  \bibinfo{author}{\bibfnamefont{B.~A.} \bibnamefont{Carreras}},
  \bibinfo{author}{\bibfnamefont{V.~E.} \bibnamefont{Lynch}}, \bibnamefont{and}
  \bibinfo{author}{\bibfnamefont{D.~E.} \bibnamefont{Newman}},
  \bibinfo{journal}{Chaos: An Interdisciplinary Journal of Nonlinear Science}
  \textbf{\bibinfo{volume}{17}}, \bibinfo{pages}{026103}
  (\bibinfo{year}{2007}).

\bibitem[{\citenamefont{Sch{\"a}ffer et~al.}(2018)\citenamefont{Sch{\"a}ffer,
  Witthaut, Timme, and Latora}}]{SWTL18}
\bibinfo{author}{\bibfnamefont{B.}~\bibnamefont{Sch{\"a}ffer}},
  \bibinfo{author}{\bibfnamefont{D.}~\bibnamefont{Witthaut}},
  \bibinfo{author}{\bibfnamefont{M.}~\bibnamefont{Timme}}, \bibnamefont{and}
  \bibinfo{author}{\bibfnamefont{V.}~\bibnamefont{Latora}},
  \bibinfo{journal}{Nature Communications} \textbf{\bibinfo{volume}{9}},
  \bibinfo{pages}{1975} (\bibinfo{year}{2018}).

\bibitem[{\citenamefont{{\'O}dor and Hartmann}(2018)}]{POWcikk}
\bibinfo{author}{\bibfnamefont{G.}~\bibnamefont{{\'O}dor}} \bibnamefont{and}
  \bibinfo{author}{\bibfnamefont{B.}~\bibnamefont{Hartmann}},
  \bibinfo{journal}{Physical Review E} \textbf{\bibinfo{volume}{98}},
  \bibinfo{pages}{022305} (\bibinfo{year}{2018}).

\bibitem[{\citenamefont{Alava et~al.}(2006)\citenamefont{Alava, Nukala, and
  Zapperi}}]{Alava_2006}
\bibinfo{author}{\bibfnamefont{M.~J.} \bibnamefont{Alava}},
  \bibinfo{author}{\bibfnamefont{P.~K. V.~V.} \bibnamefont{Nukala}},
  \bibnamefont{and} \bibinfo{author}{\bibfnamefont{S.}~\bibnamefont{Zapperi}},
  \bibinfo{journal}{Advances in Physics} \textbf{\bibinfo{volume}{55}},
  \bibinfo{pages}{349–476} (\bibinfo{year}{2006}), ISSN
  \bibinfo{issn}{1460-6976}.

\bibitem[{\citenamefont{Unicomb et~al.}(2018)\citenamefont{Unicomb, Iniguez,
  and Karsai}}]{Th-cont}
\bibinfo{author}{\bibfnamefont{S.}~\bibnamefont{Unicomb}},
  \bibinfo{author}{\bibfnamefont{G.}~\bibnamefont{Iniguez}}, \bibnamefont{and}
  \bibinfo{author}{\bibfnamefont{M.}~\bibnamefont{Karsai}},
  \bibinfo{journal}{Scientic Reports} \textbf{\bibinfo{volume}{8}},
  \bibinfo{pages}{3094} (\bibinfo{year}{2018}).

\bibitem[{\citenamefont{Mu\~{n}oz et~al.}(2010)\citenamefont{Mu\~{n}oz,
  Juh\'{a}sz, Castellano, and \'{O}dor}}]{Munoz2010}
\bibinfo{author}{\bibfnamefont{M.~A.} \bibnamefont{Mu\~{n}oz}},
  \bibinfo{author}{\bibfnamefont{R.}~\bibnamefont{Juh\'{a}sz}},
  \bibinfo{author}{\bibfnamefont{C.}~\bibnamefont{Castellano}},
  \bibnamefont{and} \bibinfo{author}{\bibfnamefont{G.}~\bibnamefont{\'{O}dor}},
  \bibinfo{journal}{Physical Review Letters} \textbf{\bibinfo{volume}{105}},
  \bibinfo{pages}{128701} (\bibinfo{year}{2010}).

\bibitem[{\citenamefont{{\'O}dor et~al.}(2015)\citenamefont{{\'O}dor, Dickman,
  and {\'O}dor}}]{HMNcikk}
\bibinfo{author}{\bibfnamefont{G.}~\bibnamefont{{\'O}dor}},
  \bibinfo{author}{\bibfnamefont{R.}~\bibnamefont{Dickman}}, \bibnamefont{and}
  \bibinfo{author}{\bibfnamefont{G.}~\bibnamefont{{\'O}dor}},
  \bibinfo{journal}{Scientific Reports} \textbf{\bibinfo{volume}{5}},
  \bibinfo{pages}{14451} (\bibinfo{year}{2015}).

\bibitem[{\citenamefont{Cota et~al.}(2018)\citenamefont{Cota, {\'O}dor, and
  Ferreira}}]{Cota_2018}
\bibinfo{author}{\bibfnamefont{W.}~\bibnamefont{Cota}},
  \bibinfo{author}{\bibfnamefont{G.}~\bibnamefont{{\'O}dor}}, \bibnamefont{and}
  \bibinfo{author}{\bibfnamefont{S.~C.} \bibnamefont{Ferreira}},
  \bibinfo{journal}{Scientific Reports} \textbf{\bibinfo{volume}{8}},
  \bibinfo{pages}{9144} (\bibinfo{year}{2018}), ISSN \bibinfo{issn}{2045-2322}.

\bibitem[{\citenamefont{Gastner and {\'O}dor}(2016)}]{CCcikk}
\bibinfo{author}{\bibfnamefont{M.}~\bibnamefont{Gastner}} \bibnamefont{and}
  \bibinfo{author}{\bibfnamefont{G.}~\bibnamefont{{\'O}dor}},
  \bibinfo{journal}{Scientific Reports} \textbf{\bibinfo{volume}{6}},
  \bibinfo{pages}{27249} (\bibinfo{year}{2016}).

\bibitem[{\citenamefont{{Di Santo} et~al.}(2018)\citenamefont{{Di Santo},
  Villegas, Burioni, and Mu{\~n}oz}}]{MunPNAS}
\bibinfo{author}{\bibfnamefont{S.}~\bibnamefont{{Di Santo}}},
  \bibinfo{author}{\bibfnamefont{P.}~\bibnamefont{Villegas}},
  \bibinfo{author}{\bibfnamefont{R.}~\bibnamefont{Burioni}}, \bibnamefont{and}
  \bibinfo{author}{\bibfnamefont{M.}~\bibnamefont{Mu{\~n}oz}},
  \bibinfo{journal}{Proceedings of the National Academy of Sciences of the
  United States of America} \textbf{\bibinfo{volume}{115}},
  \bibinfo{pages}{E1356} (\bibinfo{year}{2018}).

\bibitem[{\citenamefont{\'Odor}(2003)}]{PhysRevE.67.056114}
\bibinfo{author}{\bibfnamefont{G.}~\bibnamefont{\'Odor}},
  \bibinfo{journal}{Physical Review E} \textbf{\bibinfo{volume}{67}},
  \bibinfo{pages}{056114} (\bibinfo{year}{2003}).

\bibitem[{\citenamefont{Choi et~al.}(2018)\citenamefont{Choi, Lee, Kert\'esz,
  and Kahng}}]{PhysRevE.98.012311}
\bibinfo{author}{\bibfnamefont{W.}~\bibnamefont{Choi}},
  \bibinfo{author}{\bibfnamefont{D.}~\bibnamefont{Lee}},
  \bibinfo{author}{\bibfnamefont{J.}~\bibnamefont{Kert\'esz}},
  \bibnamefont{and} \bibinfo{author}{\bibfnamefont{B.}~\bibnamefont{Kahng}},
  \bibinfo{journal}{Physical Review E} \textbf{\bibinfo{volume}{98}},
  \bibinfo{pages}{012311} (\bibinfo{year}{2018}).

\bibitem[{\citenamefont{Harris}(1974)}]{Harris74}
\bibinfo{author}{\bibfnamefont{T.~E.} \bibnamefont{Harris}},
  \bibinfo{journal}{The Annals of Probability} \textbf{\bibinfo{volume}{2}},
  \bibinfo{pages}{969} (\bibinfo{year}{1974}).

\bibitem[{\citenamefont{Kaiser and Hilgetag}(2010)}]{KH}
\bibinfo{author}{\bibfnamefont{M.}~\bibnamefont{Kaiser}} \bibnamefont{and}
  \bibinfo{author}{\bibfnamefont{C.}~\bibnamefont{Hilgetag}},
  \bibinfo{journal}{Frontiers in Neuroinformatics} \textbf{\bibinfo{volume}{4}}
  (\bibinfo{year}{2010}).

\bibitem[{\citenamefont{Latham and S.}(2004)}]{LatNir2004}
\bibinfo{author}{\bibfnamefont{P.~E.} \bibnamefont{Latham}} \bibnamefont{and}
  \bibinfo{author}{\bibfnamefont{N.}~\bibnamefont{S.}},
  \bibinfo{journal}{Neural Compuping} \textbf{\bibinfo{volume}{16}},
  \bibinfo{pages}{1385} (\bibinfo{year}{2004}).

\bibitem[{\citenamefont{Barthelemy}(2018)}]{Barthelemy}
\bibinfo{author}{\bibfnamefont{M.}~\bibnamefont{Barthelemy}},
  \bibinfo{journal}{Comptes Rendus Physique} \textbf{\bibinfo{volume}{19}},
  \bibinfo{pages}{205–232} (\bibinfo{year}{2018}), ISSN
  \bibinfo{issn}{1631-0705}.

\bibitem[{\citenamefont{Watts and Strogatz}(1998)}]{WS98}
\bibinfo{author}{\bibfnamefont{D.~J.} \bibnamefont{Watts}} \bibnamefont{and}
  \bibinfo{author}{\bibfnamefont{S.~H.} \bibnamefont{Strogatz}},
  \bibinfo{journal}{Nature} \textbf{\bibinfo{volume}{393}},
  \bibinfo{pages}{440} (\bibinfo{year}{1998}), ISSN \bibinfo{issn}{1476-4687}.

\bibitem[{\citenamefont{Fronczak et~al.}(2004)\citenamefont{Fronczak, Fronczak,
  and Ho\l{}yst}}]{Fron}
\bibinfo{author}{\bibfnamefont{A.}~\bibnamefont{Fronczak}},
  \bibinfo{author}{\bibfnamefont{P.}~\bibnamefont{Fronczak}}, \bibnamefont{and}
  \bibinfo{author}{\bibfnamefont{J.~A.} \bibnamefont{Ho\l{}yst}},
  \bibinfo{journal}{Physical Review E} \textbf{\bibinfo{volume}{70}},
  \bibinfo{pages}{056110} (\bibinfo{year}{2004}).

\bibitem[{\citenamefont{Humphries and Gurney}(2008)}]{HumphriesGurney08}
\bibinfo{author}{\bibfnamefont{M.~D.} \bibnamefont{Humphries}}
  \bibnamefont{and} \bibinfo{author}{\bibfnamefont{K.}~\bibnamefont{Gurney}},
  \bibinfo{journal}{PLOS ONE} \textbf{\bibinfo{volume}{3}}, \bibinfo{pages}{1}
  (\bibinfo{year}{2008}).

\bibitem[{\citenamefont{Palva et~al.}(2013)\citenamefont{Palva, Zhigalov,
  Hirvonen, Korhonen, Linkenkaer-Hansen, and Palva}}]{brainexp}
\bibinfo{author}{\bibfnamefont{J.}~\bibnamefont{Palva}},
  \bibinfo{author}{\bibfnamefont{A.}~\bibnamefont{Zhigalov}},
  \bibinfo{author}{\bibfnamefont{J.}~\bibnamefont{Hirvonen}},
  \bibinfo{author}{\bibfnamefont{O.}~\bibnamefont{Korhonen}},
  \bibinfo{author}{\bibfnamefont{K.}~\bibnamefont{Linkenkaer-Hansen}},
  \bibnamefont{and} \bibinfo{author}{\bibfnamefont{S.}~\bibnamefont{Palva}},
  \bibinfo{journal}{Proceedings of the National Academy of Sciences of the
  United States of America} \textbf{\bibinfo{volume}{110}},
  \bibinfo{pages}{3585} (\bibinfo{year}{2013}).

\bibitem[{\citenamefont{Fontenele et~al.}(2019)\citenamefont{Fontenele,
  de~Vasconcelos, Feliciano, Aguiar, Soares-Cunha, Coimbra, {Dalla Porta},
  Ribeiro, Rodrigues, Sousa et~al.}}]{CC18}
\bibinfo{author}{\bibfnamefont{A.~J.} \bibnamefont{Fontenele}},
  \bibinfo{author}{\bibfnamefont{N.~A.~P.} \bibnamefont{de~Vasconcelos}},
  \bibinfo{author}{\bibfnamefont{T.}~\bibnamefont{Feliciano}},
  \bibinfo{author}{\bibfnamefont{L.~A.~A.} \bibnamefont{Aguiar}},
  \bibinfo{author}{\bibfnamefont{C.}~\bibnamefont{Soares-Cunha}},
  \bibinfo{author}{\bibfnamefont{B.}~\bibnamefont{Coimbra}},
  \bibinfo{author}{\bibfnamefont{L.}~\bibnamefont{{Dalla Porta}}},
  \bibinfo{author}{\bibfnamefont{S.}~\bibnamefont{Ribeiro}},
  \bibinfo{author}{\bibfnamefont{A.~J.~a.} \bibnamefont{Rodrigues}},
  \bibinfo{author}{\bibfnamefont{N.}~\bibnamefont{Sousa}},
  \bibnamefont{et~al.}, \bibinfo{journal}{Physical Review Letters}
  \textbf{\bibinfo{volume}{122}}, \bibinfo{pages}{208101}
  (\bibinfo{year}{2019}).

\bibitem[{\citenamefont{Moretti and Mu\~{n}oz}(2013)}]{Moretti2013}
\bibinfo{author}{\bibfnamefont{P.}~\bibnamefont{Moretti}} \bibnamefont{and}
  \bibinfo{author}{\bibfnamefont{M.~A.} \bibnamefont{Mu\~{n}oz}},
  \bibinfo{journal}{Nature Communications} \textbf{\bibinfo{volume}{4}},
  \bibinfo{pages}{2521} (\bibinfo{year}{2013}).

\bibitem[{\citenamefont{{\'O}dor}(2016)}]{CCdyncikk}
\bibinfo{author}{\bibfnamefont{G.}~\bibnamefont{{\'O}dor}},
  \bibinfo{journal}{Physical Review E} \textbf{\bibinfo{volume}{94}},
  \bibinfo{pages}{062411} (\bibinfo{year}{2016}).

\bibitem[{\citenamefont{{\'O}dor}(2019)}]{CC-tdepcikk}
\bibinfo{author}{\bibfnamefont{G.}~\bibnamefont{{\'O}dor}},
  \bibinfo{journal}{Physical Review E} \textbf{\bibinfo{volume}{99}},
  \bibinfo{pages}{012113} (\bibinfo{year}{2019}).

\bibitem[{\citenamefont{Moosavi et~al.}(2017)\citenamefont{Moosavi, Montakhab,
  and Valizadeh}}]{Afshin}
\bibinfo{author}{\bibfnamefont{S.}~\bibnamefont{Moosavi}},
  \bibinfo{author}{\bibfnamefont{A.}~\bibnamefont{Montakhab}},
  \bibnamefont{and}
  \bibinfo{author}{\bibfnamefont{A.}~\bibnamefont{Valizadeh}},
  \bibinfo{journal}{Scientific Reports} \textbf{\bibinfo{volume}{7}},
  \bibinfo{pages}{7107} (\bibinfo{year}{2017}).

\bibitem[{\citenamefont{Zarepour et~al.}(2019)\citenamefont{Zarepour, Perotti,
  Billoni, Chialvo, and Cannas}}]{PhysRevE.100.052138}
\bibinfo{author}{\bibfnamefont{M.}~\bibnamefont{Zarepour}},
  \bibinfo{author}{\bibfnamefont{J.~I.} \bibnamefont{Perotti}},
  \bibinfo{author}{\bibfnamefont{O.~V.} \bibnamefont{Billoni}},
  \bibinfo{author}{\bibfnamefont{D.~R.} \bibnamefont{Chialvo}},
  \bibnamefont{and} \bibinfo{author}{\bibfnamefont{S.~A.}
  \bibnamefont{Cannas}}, \bibinfo{journal}{Physical Review E}
  \textbf{\bibinfo{volume}{100}}, \bibinfo{pages}{052138}
  (\bibinfo{year}{2019}).

\bibitem[{\citenamefont{Mu{\~n}oz et~al.}(1999)\citenamefont{Mu{\~n}oz,
  Dickman, Vespignani, and Zapperi}}]{Munoz99}
\bibinfo{author}{\bibfnamefont{M.~A.} \bibnamefont{Mu{\~n}oz}},
  \bibinfo{author}{\bibfnamefont{R.}~\bibnamefont{Dickman}},
  \bibinfo{author}{\bibfnamefont{A.}~\bibnamefont{Vespignani}},
  \bibnamefont{and} \bibinfo{author}{\bibfnamefont{S.}~\bibnamefont{Zapperi}},
  \bibinfo{journal}{Physical Review E} \textbf{\bibinfo{volume}{59}},
  \bibinfo{pages}{6175} (\bibinfo{year}{1999}).

\bibitem[{\citenamefont{Burioni and Cassi}(1996)}]{PhysRevLett.76.1091}
\bibinfo{author}{\bibfnamefont{R.}~\bibnamefont{Burioni}} \bibnamefont{and}
  \bibinfo{author}{\bibfnamefont{D.}~\bibnamefont{Cassi}},
  \bibinfo{journal}{Physical Review Letter} \textbf{\bibinfo{volume}{76}},
  \bibinfo{pages}{1091} (\bibinfo{year}{1996}).

\bibitem[{\citenamefont{Burioni et~al.}(2000)\citenamefont{Burioni, Cassi, and
  Vezzani}}]{Burioni_2000}
\bibinfo{author}{\bibfnamefont{R.}~\bibnamefont{Burioni}},
  \bibinfo{author}{\bibfnamefont{D.}~\bibnamefont{Cassi}}, \bibnamefont{and}
  \bibinfo{author}{\bibfnamefont{A.}~\bibnamefont{Vezzani}},
  \bibinfo{journal}{The European Physical Journal B}
  \textbf{\bibinfo{volume}{15}}, \bibinfo{pages}{665–672}
  (\bibinfo{year}{2000}), ISSN \bibinfo{issn}{1434-6028}.

\bibitem[{\citenamefont{Burioni and Cassi}(2005)}]{Burioni_2005}
\bibinfo{author}{\bibfnamefont{R.}~\bibnamefont{Burioni}} \bibnamefont{and}
  \bibinfo{author}{\bibfnamefont{D.}~\bibnamefont{Cassi}},
  \bibinfo{journal}{Journal of Physics A: Mathematical and General}
  \textbf{\bibinfo{volume}{38}}, \bibinfo{pages}{R45} (\bibinfo{year}{2005}).

\bibitem[{\citenamefont{Mill\'an et~al.}(2019)\citenamefont{Mill\'an, Torres,
  and Bianconi}}]{PhysRevE.99.022307}
\bibinfo{author}{\bibfnamefont{A.~P.} \bibnamefont{Mill\'an}},
  \bibinfo{author}{\bibfnamefont{J.~J.} \bibnamefont{Torres}},
  \bibnamefont{and} \bibinfo{author}{\bibfnamefont{G.}~\bibnamefont{Bianconi}},
  \bibinfo{journal}{Physical Review E} \textbf{\bibinfo{volume}{99}},
  \bibinfo{pages}{022307} (\bibinfo{year}{2019}).

\bibitem[{\citenamefont{Esfandiary et~al.}(2020)\citenamefont{Esfandiary,
  Safari, Renner, Moretti, and Mu\~noz}}]{PhysRevResearch.2.043291}
\bibinfo{author}{\bibfnamefont{S.}~\bibnamefont{Esfandiary}},
  \bibinfo{author}{\bibfnamefont{A.}~\bibnamefont{Safari}},
  \bibinfo{author}{\bibfnamefont{J.}~\bibnamefont{Renner}},
  \bibinfo{author}{\bibfnamefont{P.}~\bibnamefont{Moretti}}, \bibnamefont{and}
  \bibinfo{author}{\bibfnamefont{M.~A.} \bibnamefont{Mu\~noz}},
  \bibinfo{journal}{Physical Review Research} \textbf{\bibinfo{volume}{2}},
  \bibinfo{pages}{043291} (\bibinfo{year}{2020}).

\end{thebibliography}

\end{document}